\documentclass[12pt]{article}
\usepackage[english]{babel}
\usepackage{dirtytalk}
\usepackage[utf8x]{inputenc}
\pdfoutput=1
\usepackage{arxiv}

\usepackage{setspace}
\usepackage{apacite}
\usepackage{url}
\usepackage{hyperref}
\usepackage{filecontents}
\usepackage{graphics}
\usepackage{pdfpages}
\usepackage{amssymb,amsmath,amsfonts,eurosym,geometry,ulem,graphicx,caption,color,setspace,sectsty,comment,natbib, footmisc,caption,pdflscape,subfigure,array,hyperref}

\newcolumntype{L}[1]{>{\raggedright\let\newline\\arraybackslash\hspace{0pt}}m{#1}}
\newcolumntype{C}[1]{>{\centering\let\newline\\arraybackslash\hspace{0pt}}m{#1}}
\newcolumntype{R}[1]{>{\raggedleft\let\newline\\arraybackslash\hspace{0pt}}m{#1}}

\begin{document}

\begin{titlepage}
\title{What Explains Gender Gap in Unpaid Household and Care Work in India?}
\author {Athary Janiso\thanks{Research Scholar, Department of Economics and  Finance, Birla Institute of Technology and Science-Pilani, Hyderabad campus, E-mail: p20190026@hyderabad.bits-pilani.ac.in}\and Prakash Kumar Shukla  \thanks{Research Scholar, Department of Economics and  Finance, Birla Institute of Technology and Science-Pilani, Hyderabad campus, E-mail: p20190027@hyderabad.bits-pilani.ac.in} \and  Bheemeshwar Reddy A\thanks{Assistant Professor, Department of Economics and  Finance, Birla Institute of Technology and Science-Pilani, Hyderabad campus, E-mail: bheem@hyderabad.bits-pilani.ac.in }}
\date{}
\maketitle
\begin{abstract}
Due to the unavailability of nationally representative data on time use, a systematic analysis of the gender gap in unpaid household and care work has not been undertaken in the context of India. The present paper, using the recent Time Use Survey (2019) data, examines the socioeconomic and demographic factors associated with variation in time spent on unpaid household and care work among men and women. It analyses how much of the gender gap in the time allocated to unpaid work can be explained by differences in these factors. The findings show that women spend much higher time compared to men in unpaid household and care work. The decomposition results reveal that differences in socioeconomic and demographic factors between men and women do not explain most of the gender gap in unpaid household work. Our results indicate that unobserved gender norms and practices most crucially govern the allocation of unpaid work within Indian households.


%\bigskip
\end{abstract}
\setcounter{page}{0}
\thispagestyle{empty}
\end{titlepage}
%\pagebreak \newpage

\section{Introduction} \label{sec:Introduction}
Women continue to bear the unequal burden of unpaid household and care work both in developed and developing countries \citep{united2018time-use}. Such intra-household uneven distribution of unpaid work can be detrimental to women's wellbeing . The excess load of unpaid household and care responsibilities for women not only deters their access to paid work, lowering their educational attainment and earnings, but also causes distress to their health \citep{hirway2009unpaid, hirsch2013effect, fendel2020effect}. For example, a recent study from China points out that time devoted to unpaid work at home at the expense of sleep and leisure can have negative consequences for women's mental health \citep{liu2018gender}. Moreover, when women participate in paid work, the inequitable gender distribution of unpaid work at home often deprives them of adequate time for self-care and leisure activities \citep{dong2015gender, chopra2017no}. Given the many faceted consequences for women’s wellbeing, it is important to investigate the gender gap in unpaid household and care work and the factors that explain this gap.

In South Asia, unpaid housework is still predominantly a woman’s responsibility \citep{hirway2015unpaid}. Though large section of women are engaged in back breaking work in sectors such as agriculture and construction in India, they also shoulder most of the unpaid household work. This is true across India, irrespective of the regional variations in the social and cultural norms and economic development \citep{ shimray2004women, luke2014husbands, lahiri2014women, dutta2016extent, manhas2017analysis, irani2020getting,swaminathan2020women,rao2021work}. Recent improvements in women’s educational attainment in India should have ideally decreased women’s burden of unpaid housework in accordance with the relative resources theory \citep{blood1960husbands}.  The rise in educational attainment did not increase women’s participation in paid work in India. On the contrary, women’s labour force participation rate has been declining despite the rapid economic growth\footnote{According to National Statistical Office (NSO) Periodic Labour Force Survey (PLFS) (2017-18), the female labour force participation rate decreases from 29.4 per cent to 17.5 per cent between 2004-05 to 2017-18, while it remains steady for men at 55 per cent during the same period.}. An increasing proportion of women in India are hence attending to domestic duties due to various demand and supply-side factors\citep{Naidu, afridi2018fewer}. In the context of these two significant changes in the educational and employment status of women, it is imperative to study the unequal distribution of unpaid household and care work in India. 

Most of the empirical work on the factors explaining the gender gap in unpaid work within households draws on research from developed countries. However, some recent work has emerged from other contexts such as China, Kyrgyzstan and Latin American countries \citep{walker2013time, luke2014husbands, dong2015gender, amarante2018unfolding, kolpashnikova2020gender, balde2020unpaid, torabi2020spouses, de2020urbanization, dominguez2021gender}. A systematic analysis of the gender gap in unpaid household and care work has not been undertaken in the Indian context primarily due to the lack of time use survey data at the national level. The recent nationally representative Time Use Survey (2019) (henceforth, TUS-2019) allows us to examine the Indian case comprehensively. This paper addresses two main concerns. First, it investigates the association between time spent on unpaid household work and care work by women and men per day and their individual and household level socioeconomic and demographic characteristics. Second, it examines to what extent the gender gap in unpaid household and care work is owing to the differences in socioeconomic and demographic factors between women and men in India. 

This paper contributes to the existing literature on gender inequality in time devoted to unpaid work at home in the following ways. To the best of our knowledge this is the first attempt to provide evidence on the factors associated with individual time allocation in unpaid household and care work using the nationally representative data from TUS-2019 in the context of India. In the absence of TUS, previous studies had limited sample size and hence their findings cannot be generalised to the all India level.  Further, this is the first study to explain the gender gap in unpaid work in India by employing the Oaxaca–Blinder decomposition analysis. Our paper differs from other existing studies as it addresses some of methodological concerns related to OLS estimates: we check for robustness of our results by using household fixed effects, Tobit and Seemingly Unrelated Regression (SUR) specification. Finally, one of the crucial determinants of gender gap in unpaid household and care work is the wages of individuals\citep{kan2008does}. However,  TUS-2019 does not contain wage data. This paper overcomes this hurdle by incorporating predicted wages, generated by using periodic labour force survey 2018-19, as an additional covariate in our analysis.

Our results confirm a large gap in time spent on unpaid household work between women and men after accounting for different socioeconomic and demographic factors. The decomposition results show that majority of the gender gap in unpaid work cannot be attributed to differences in endowments between women and men. In other words, most of the gap in unpaid work at home between men and women in India can be chalked up to unobserved factors such as gender norms and practices. Our main findings are consistent with most of the evidence found in context of developing countries. 
\section{Theoretical perspectives on gender gap in unpaid work}\label{sec:Theoretical }
Both economists and sociologists have provided theoretical and empirical explanations for unequal distribution of unpaid work between men and women within households. Broadly two theoretical frameworks have emerged: bargaining theory \citep{mcelroy1990empirical, hersch1994housework, lundberg1996bargaining} or relative resources theory \citep{blood1960husbands, greenstein2000economic,  gupta2008whose, baxter2013negotiating} and gender based theory. From a bargaining theory perspective, individuals’ relative bargaining power within households is a key determinant of intra-household time allocation. Within this framework, bargaining between partners is motivated by the desire to prevent the breakdown of marriage. Individuals’ resources such as earnings from market or assets determine their bargaining position within the households. Therefore, the partner with the greater resources will be able to negotiate spending less time on unpaid housework. 

However, bargaining models do not account for the differences in the gender of the household members \citep{agarwal1997bargaining, kabeer1997women, kantor2003women}. They do not adequately explain the role of social and cultural norms in limiting the bargaining ability of women within households \citep{agarwal1997bargaining}. For instance, in case of India mere access to resources within a marriage does not translate into increased bargaining power for women as they are unequal partners in marriages owing to social and cultural factors such as age of marriage and stigma attached to divorce \citep{kantor2003women}.  Hence, the bargaining models do not fully explain the uneven distribution of unpaid work between women and men in socio-cultural contexts different from the Global North from where these theories have primarily emerged. 

The gender-based approaches argue that gender norms and ideology crucially determine the gender gap in unpaid work within households. Scholars have highlighted different aspects of gender within this broad framework. Women are disproportionately burdened with unpaid household work owing to their gender identity and prevailing societal norms dictating gender roles in a particular social context. This is explained through two kinds of gendered behaviour. First, women through performance of household tasks and men through their non-performance of household tasks conform their identities as feminine and masculine respectively. Hence, housework is a medium of “doing” gender \citep{berk1985relationship, ferree1990beyond}. Second, household work becomes an arena to re-affirm gender roles which have been upturned for reasons like equal or more income of women and men’s unemployed status. Hence, contrary to the relative resources theory, women’s higher income does not lead to their higher bargaining power. Rather, women engage in more housework to establish their femininity and men withdraw from it to compensate for their inability to perform the socially expected role of the breadwinner \citep{brines1994economic,   kan2008does, sevilla2010gender}.

\section{Factors affecting time allocation to unpaid work } \label{sec:Factors affecting  }
Factors associated with an individual's time allocation to unpaid work at home can be broadly grouped into three categories — individual characteristics, household characteristics and institutional factors. Some of the important individual factors include marital status, educational attainment, employment, and wages. A number of studies from Australia, Italy, United States suggest that married women do more housework than other women \citep{shelton1993does, south1994housework, gupta1999effects,  baxter2005marry, meggiolaro2014household},while no such difference was found in case of married men \citep{shelton1993does, south1994housework}. 

In the existing  literature education has been seen as affecting gender parity in household work through two channels. Higher educational attainment of women raises their relative resources, reducing time spent on unpaid household work \citep{sullivan2018gendered}. Second, individuals with higher education are more likely to subscribe to more egalitarian gender views, resulting in more equitable sharing of unpaid household work \citep{altintas2016fifty}. Thus the cumulative effect of increasing education is that women do less housework and men participate more in unpaid household work \citep{sullivan2018gendered}. Evidence from different countries suggests that women's educational attainment has a negative relationship with the amount of time they spend on unpaid household work \citep{shelton1992women, brines1994economic, torabi2020spouses}. \cite{kolpashnikova2021educational}, empirically examined the relationship between educational attainment and time spent on unpaid household work among Japanese, Taiwanese, and American women. They find that while the negative relationship between educational attainment and time spent on household work was corroborated in the case of single women in all countries under study, the negative relationship was not true for married women with children in Taiwan and married Japanese women. 

In developed countries, women employed in paid work do less unpaid work compared to unemployed women. For example, \cite{van2018unemployment}, using European Social Survey data from 27 countries, finds that unemployed men and women spend more time on unpaid household work than those who are employed. However, they note that unemployed women do more additional household work than unemployed men. A study based in the United States shows that irrespective of the partner’s or the family’s total earnings, time spent by women in housework goes down with rise in their earnings \citep{gupta2008whose}.

Household characteristics impacting gender parity in unpaid housework are time-saving domestic equipment, outsourcing household work and the number of children or additional adults in the household. For example, \cite{gershuny1988historical} argue that the reduction in the time spent on household work by women in the UK and USA could be attributable to the introduction of the time-saving features of new household appliances such as dishwashers and microwave oven. \cite{kizilirmak2009unequal} using data from TUS carried out in South Africa find that lower income increased the amount of time devoted to unpaid work by women, but  it did not affect men's unpaid work time. Evidence suggests that the presence of additional adult and adolescent in the household reduced the time devoted to unpaid housework and care work for all individuals \citep{cheal2003children, lee2003children, kalenkoski2011time}. However, \cite{gershuny2014household} found that in the UK the effect of this factor differed by gender, and women's time spent on housework did not decrease. There is compelling evidence to suggest that additional number of children in the household increases the care work and household work for women \citep{ neilson2014s,  torabi2020spouses}. The type of household to which women belong is a critical predictor of time spent on unpaid work. \cite{srivastava2020time} argues that the division of unpaid household work differs substantially between multigenerational households and nuclear households in India. 

Finally, some of the institutional factors at national level that may impact the distribution of unpaid work within households include female labour force participation, gender norms, welfare regimes and the level of economic development \citep{fuwa2004macro, hook2006care, anxo2011gender, campana2017increasing, amarante2018unfolding, dominguez2021gender}. In a comparative study of Mexico, Peru, and Ecuador, \cite{campana2017increasing} show that the gendered allocation of total work has greater levels of parity in countries with more egalitarian gender norms.

\includepdf[page={1}]{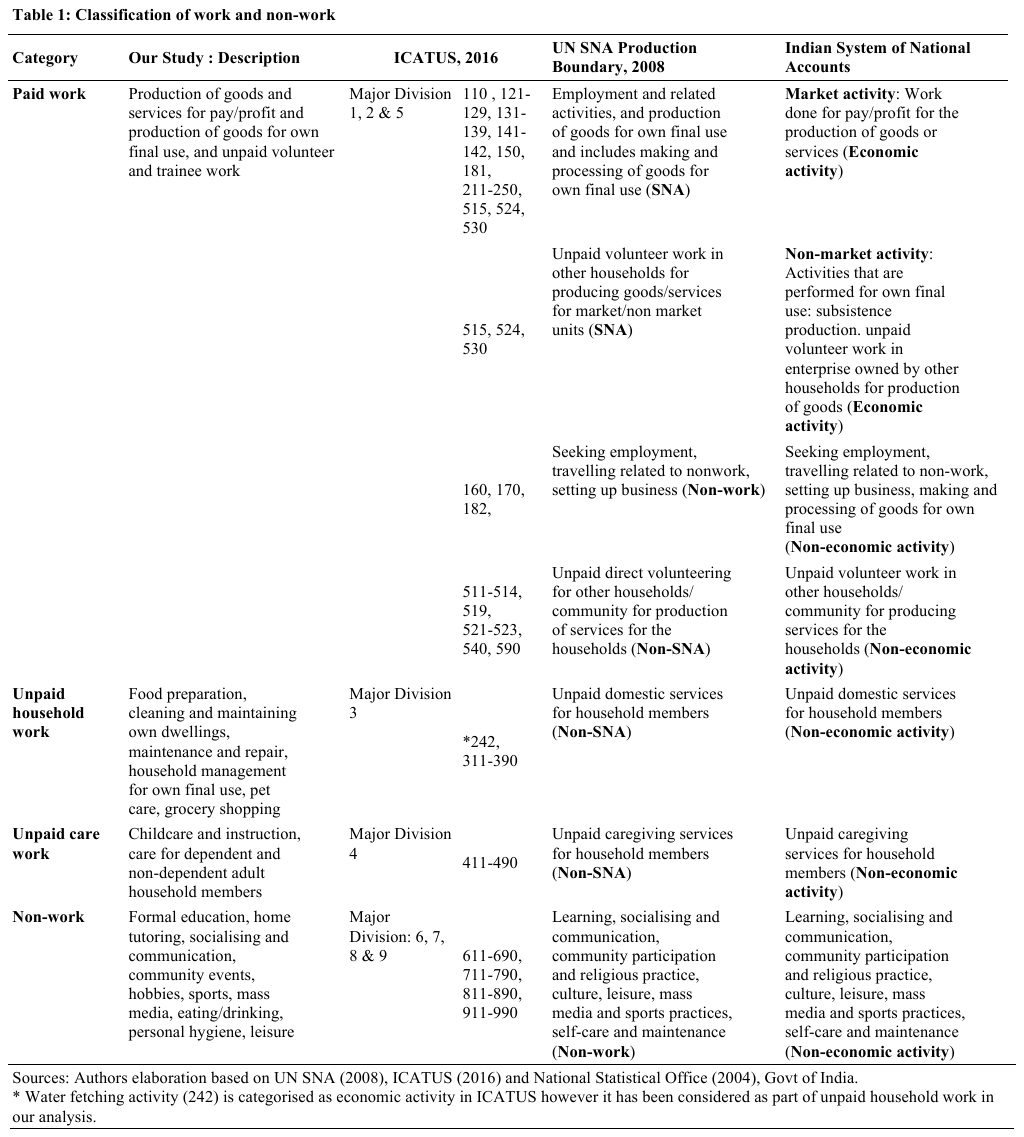}

\section{Data} \label{sec:data}
We use the first nationally representative TUS conducted by the National Statistical Office (NSO) from January 2019 to December 2019 to examine gender-based disparity in unpaid household and care work in India\footnote{ A pilot TUS was carried out in 1997-98 in only six states in India}. The survey covered 138,799 sample households across rural and urban India. From each member of the sample household, who was above five years old, the survey collected information about how they had spent their time in the last 24 hours (04:00 am on the previous day to 4:00 am on the day of survey) at 30 minutes interval. Additionally, it also recorded multiple and simultaneous activities done by respondents during the 30 minutes time slot. It was ascertained whether the day was a normal working day or other (non-working) day for each eligible individuals in the household. Other days may include holidays and non-working weekends for the employed person. The data set also contains information on household characteristics such as income (monthly household consumer expenditure), social group, religion, and individual characteristics such as employment status, education and other demographic information. The time spent on different activities by individuals during reference day (24 hours) was classified into nine major divisions of activities as per the International Classification of Activities for Time Use Statistics (ICATUS) \citep{united2016international}.

For this study, we classify the total time spent by an individual in a day into four broad activities: unpaid household work, unpaid care work; paid work; non-work. Especially, the advantage of dividing unpaid work at home into unpaid household work and unpaid care work is that it allows us to distinguish between the utility and social norms connected with care work and unpaid household work (Kizilirmak and Memis, 2009). Table 1 illustrates how each of these four activities map into the work and non-work definitions of the production boundary of the United Nations System of National Accounts \citep{united2009system} and the Indian Systems of National Accounts. The time spent on unpaid household work mainly comprise of activities such as food and meals management and preparation, cleaning and maintaining of own dwelling and surroundings, do-it-yourself decoration, maintenance and repair, care and maintenance of textiles and footwear, household management for own final use, pet care, shopping for own household members and fetching water from natural and other sources for own final use during the reference day. In the context of rural India, fetching water becomes crucial as it is typically carried out by women and consumes a considerable amount of time in a day for women \citep{motiram2010social}. The time spent on unpaid care work by individuals comprises of time devoted to caring services provided by them to dependent and non-dependent children and adults in their households in a day. The time spent on paid work by an individual includes the total time devoted to employment and related activities and production of goods for their final use, but excludes time spent by them to fetch water from natural and other sources for their final use in a day. Time spent in non-work includes all activities such as leisure, self-care, and learning.

 We restrict our analysis to the sample of males and females who are 15 years of age and above. In addition, our work sample does not include individuals living in single-member households. Households consisting of only male members or only female members have also been excluded from the analysis. The restrictions imposed for this analysis leaves us with 182224 males and 175191 females.

\includepdf[page={1}]{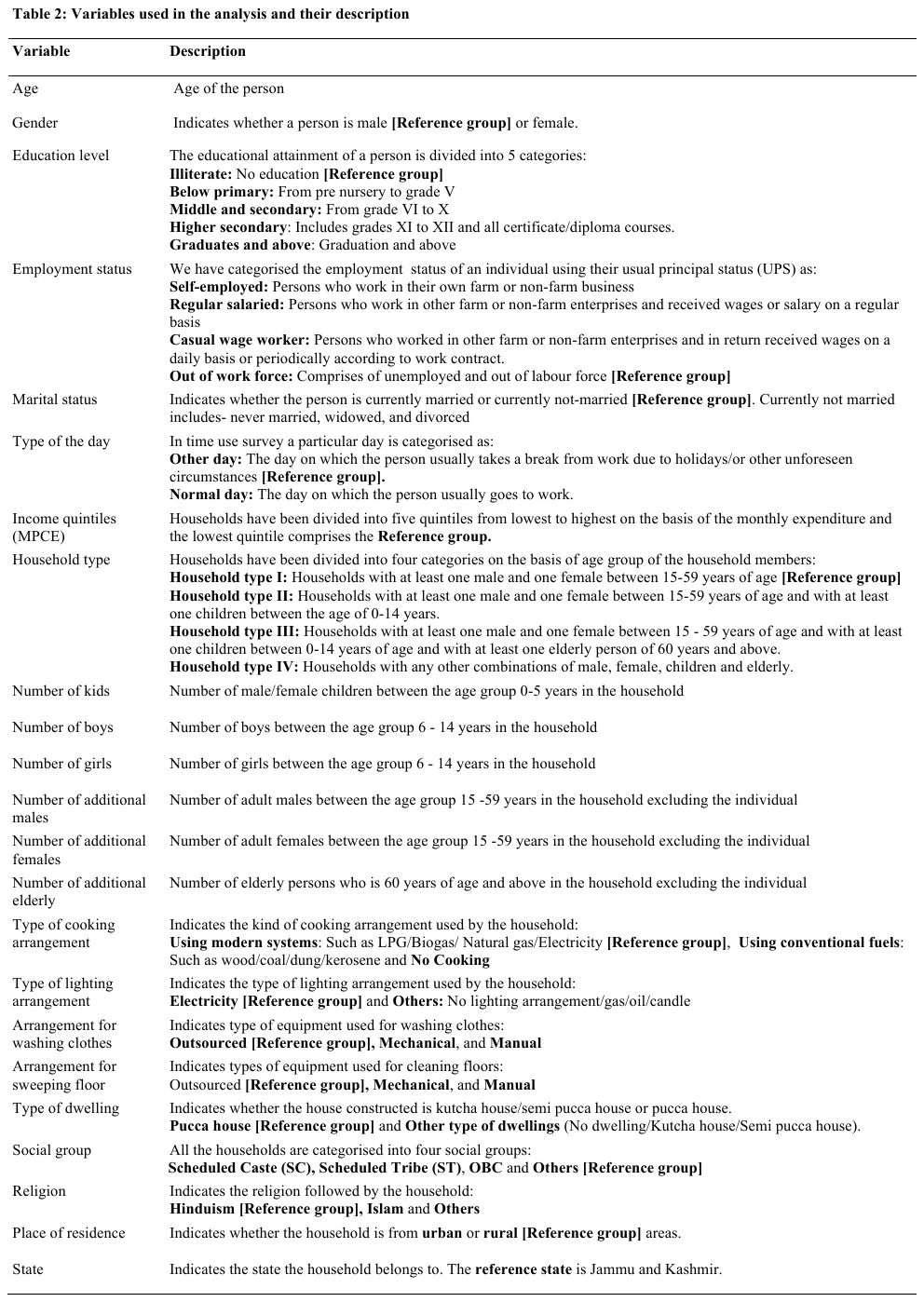}

\section{Methodology} \label{sec:Methodology}
Studies are divided on the appropriate econometric model that should be used for analysing factors associated with the amount of time devoted by individuals to unpaid household and care work using TUS data \citep{kalenkoski2011time, stewart2013tobit, foster2013tobit}. The main reason for the disagreement among researchers is that many respondents, especially men, report zero time spent in unpaid household and care activities. Thus Ordinary Least Squares (OLS) estimates yield biased and inconsistent results as OLS does not consider the censored aspects of time use data \citep{greene2003econometric}. Hence, it is argued by some scholars that Tobit is the appropriate model for taking into account the censored data \citep{kalenkoski2011time, foster2013tobit}. The justification for using the Tobit model depends on the assumption that zeros in time use data represent real non-participation in any given activity. However, \cite{stewart2013tobit} argues that if zeros in the data represent random measurement errors and capture infrequency rather than censoring, the OLS estimates are more appropriate than the Tobit estimates. It is not very easy to ascertain why zeros exist in TUS data . Given that there is no consensus among the researchers on the most apposite way to model time use data, we have estimated both OLS and Tobit regressions to examine the association between different socioeconomic and demographic characteristics of individuals and the amount of time spent by them in unpaid household and care work per day. 
Regression Eq. (1) is estimated using OLS method.        $$T_{i}= \beta_{0} +\beta_{1} X+ u_{i} \hspace{2cm}\left(1\right)$$$T_{i}$ is time spent by individual $i$ in minutes on unpaid household work(or care work) per day, and $X$ represents a vector of explanatory variables. The covariates in the regression model for women and men include individual characteristics such as age, age square, marital status, education, employment status (casual worker or regular salaried worker or self-employed or out-of-work), and type of day (normal day or other days). Other covariates include household-level characteristics such as household income (monthly per capita expenditure), type of household, the number of children aged up to 5 years and number of girls and boys aged from 6 to 14, additional number of males and females in the age group of 15-59 in the household. We also include the place of residence (rural or urban), state of residence, caste and religion of the household. Other independent variables incorporated in the model include household amenities like possession of  modern cooking facilities (LPG connection) and washing machine. Table 2 provides the complete list of variables with their definition included in the analysis, including reference categories used for categorical variables. We use the same set of dependent variables and independent variables in the Tobit specifications as in the OLS specifications.

 On a given day, an increase in time allocation to unpaid household work by an individual may imply lower time available for unpaid care work, paid work and non-work activity. In other words, individuals' decision to allocate time among the four activities is a simultaneous decision. Therefore, following \cite{neuwirth2007determinants} and \cite{dong2015gender}, we jointly estimate a set of Seemingly Unrelated Regression (SUR) equations to examine to what extent the time allocated by individuals to unpaid household work, unpaid care work, paid work, and non-work varies by socioeconomic  and demographic characteristics among women and men. The SUR yields more efficient estimates than OLS estimate if the error terms across four equations  are correlated \citep{zellner1962efficient}. To account for the interdependence of four activities, we impose two restrictions on the SUR estimation: the first restriction is the sum of the four equations' intercepts equal to 1440 minutes (i.e., 24 hours) per day and the second restriction is the sum of the coefficients of each covariate over all four activities equal to zero.  
 
We run regressions on Eq.(2) both together and separately for men and women to model the association between time spent on activities (i.e., unpaid household work,  unpaid care work, paid work and non-work activity) and the different covariates.
\\
\begin{center}
         $$T_{ji}= \beta_{j0} +\beta_{j\alpha} X_{ji}+ u_{ji},\hspace{1cm} i=1, ...,n \hspace{2cm}\left(2\right)$$\\
         \end{center}
 
   $$   \sum\limits_{j=1}^{4} \beta_{j0}=1440  \hspace{0.3cm} and \; \sum\limits_{j=1}^{4} \beta_{j\alpha}=0 \hspace{1cm}  \forall\:{\alpha= 1, 2, ... K} $$\\

We jointly estimate $j$ equation, one for each activity, indexed by $j= 1, 2, 3, 4$. Where $T_{ji}$ is time in minutes on a given activities $j$ by individual $i$ per day; $X_{ji}$ is a vector of covariates capturing individuals' socioeconomic and demographic characteristics. $\beta_{j0}$ represents intercept; $\beta_{j\alpha}$ is a vector of slope coefficients that indicates the impact of covariates on  time devoted  to an activity  $j $ per day and $u_{ji}$ represents the error term. 

Our  next objective is to examine to what extent  the male-female gap in time devoted to unpaid work is on account of gender differences in socioeconomic and demographic characteristics. We have employed Oaxaca-Blinder decomposition method for this purpose \citep{blinder1973wage, oaxaca1973male}\footnote{ We utilise the “twofold” decomposition that is commonly used in the wage discrimination studies \citep{jann2008blinder}.Among others \cite{amarante2018unfolding}, and \cite{kolpashnikova2020gender} have recently employed Oaxaca–Blinder decomposition  method to explain gender gap in unpaid work in the context of Latin America and Kyrgyzstan respectively}. 

$$ \bar{T^{w}}-\bar{T^{m}} =\left[ \left(\bar{X_{w}}-\bar{X_{m}}\right)\hat{\beta_{w}} \right]+ \left[\bar{X_{m}} \left(\hat{\beta_{w}}-\hat{\beta_{m}}\right) \right] \hspace{2cm}\left(3\right)  $$
The advantage of using Oaxaca-Blinder method is that it allows us to decompose the average gap between men and women in time spent in unpaid work, $ \bar{T^{w}}-\bar{T^{m}}$,   into two components: "endowment effect" (or "explained part") and the "coefficient effect" (or "unexplained" part). The first term in Eq. (3) represents the "explained" gap, and this gap is attributed to differences in the distribution of socioeconomic and demographic factors between women and men. "Explained" gap reflects a counterfactual comparison of gap in time devoted to unpaid work if women had the same characteristics as men. The second term in Eq. (3) denotes the "unexplained" gap. The "unexplained" gap cannot be explained by differences in the distribution of characteristics of women and men. Instead, the "unexplained" gap is due to the different effect of covariates for males and females which are captured by the coefficients. Differences in the responses of covariates between women  and men may arise due to unobserved factors. The "unexplained" part is often attributed to discrimination or the gender-based social norms around unpaid household and  care work and other relevant factors that are not included in the model\footnote{The Blinder-Oaxaca decomposition has the drawback that the findings are dependent on the categorical predictors' reference category. To get around this issue, we employ the normalisation option \citep{oaxaca1999identification}}.

In addition to the above standard Oaxaca-Blinder decomposition, for robustness check, we also estimate Tobit decomposition following \cite{bauer2008extension} extension of Oaxaca-Blinder decomposition for Tobit models. Given the intuitive interpretation of OLS estimates, we present the same in the results section and the Tobit and SUR estimates are provided in the robustness check section.

Our regression specifications do not account for the potential endogeneity of some of the independent variables. For example, we assume that an individual’s household per capita income is an exogenous variable.  The causal association between household per capita income and time spent doing unpaid labour, on the other hand, could go both ways. Higher household income allows individuals to outsource unpaid domestic chores and childcare.  On the contrary, an increase in time devoted to unpaid housework may reduce one’s time dedicated to paid work, lowering income \citep{noonan2001impact}. Thus, our results only emphasise the statistical association between the covariates and time spent on unpaid work, and they may not indicate a causal relationship. 

\includepdf[page={1,2}]{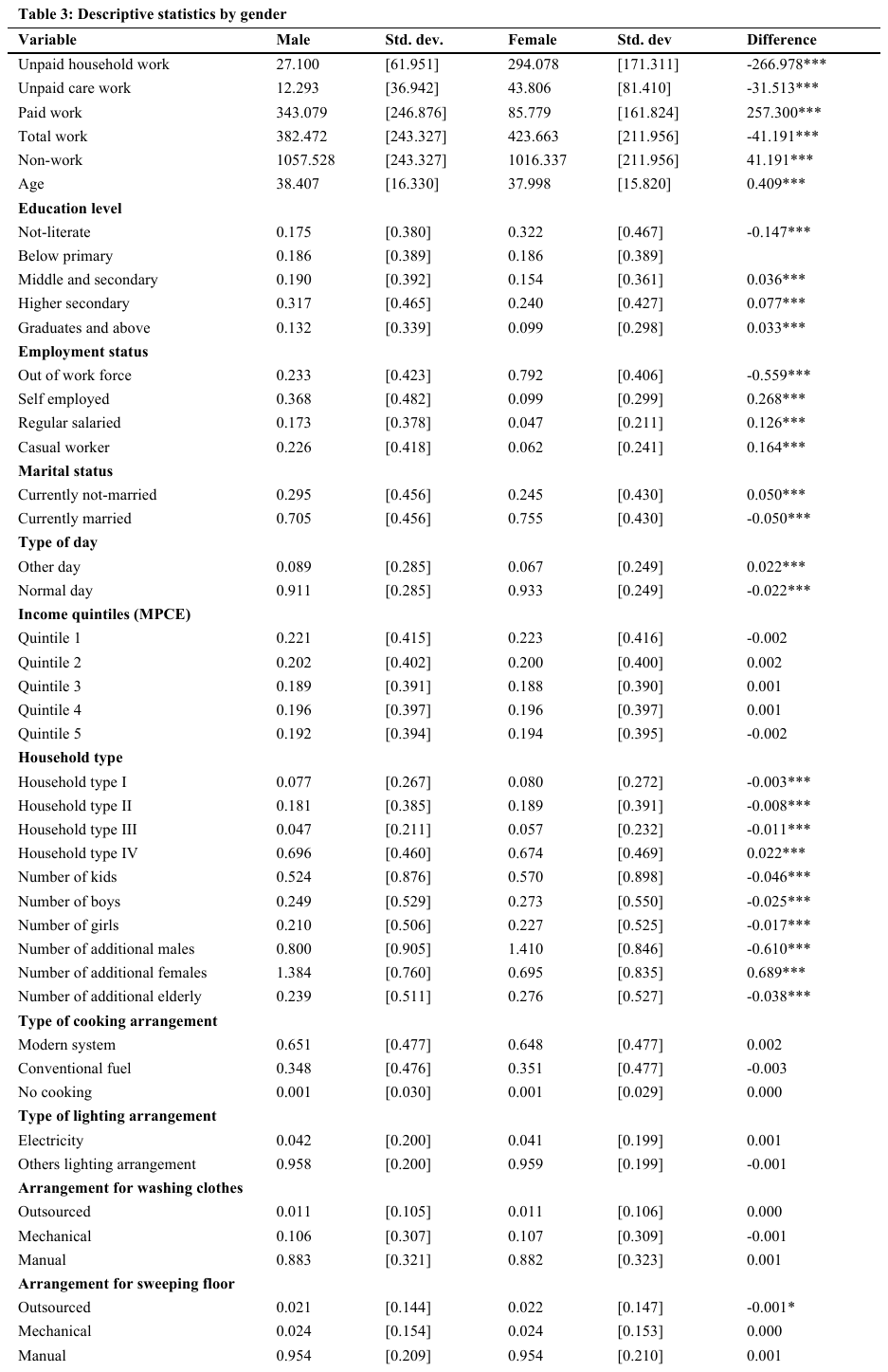}

\section{Results}{\label{sec:Results}}

\subsection{Factors associated with time spent in unpaid work}

Before discussing the relationship between different factors and time spent in unpaid work by men and women, we present the descriptive statistics by gender in Table 3.  The average time spent in unpaid household and care work by women is much higher than men. For instance, the average time spent on unpaid household work per day by women is 294 minutes, as opposed to merely 27 minutes spent by men on the same. Though men spend relatively more time in paid work, the total of women's paid and unpaid work in a day is 41 minutes more than that of men. The descriptive statistics also demonstrate that compared to women, men have better education and labour market outcomes. For instance, 32.2 percent of women are not literate compared to only 17.5 percent of men. Less than 10 percent of women have at least completed undergraduate education as against 13.2 percent of men. Similarly, there are significant differences in work participation rate of men and women: 76.7 percent of men are employed, which is significantly higher than the 20.8 percent of women. Also, women are disproportionately self-employed as compared to men.

Table 4 and Table 5 report the results of Eq. (1) OLS estimation of the association between the amount of time spent by an individual on unpaid household and care work per day and a set of socioeconomic and demographic variables. Gender dummies indicate that women spend significantly more time in unpaid household and care work than their male counterparts after accounting for other relevant covariates. For example, the estimated coefficient of gender, column 2 in Table 4, shows that women spend around three hours (i.e., 173 minutes) more than men on unpaid household work per day. Further, the specification in column 4 in Table 4 with household fixed effects to control for unobserved household-level factors also show similar results as column 2 in Table 4. Similarly, if we consider the entire sample, women work about 20 minutes more than men in unpaid care work per day. However, if we restrict the sample only to households in which at least one member spent more than zero minutes on care work, the gender dummy shows that the gap in time spent on unpaid care work between women and men increases three folds per day (see column 1 Table A1 in Appendix).

Given that gender is an essential predictor of the amount of time spent by individuals on unpaid household work, we estimated separate regression models to examine how different factors affect time spent by women and men in unpaid household work and care work.  The regression results show that married women spend around two hours (128 minutes) more than women in the reference category (unmarried or divorced or separated) in unpaid household work per day. Similarly, the gap in the amount of time devoted to unpaid care work between currently married and rest of the women is 24 minutes. Currently married men spend only marginally more on unpaid household work as compared to rest of the men. Likewise, marriage leads to a small but statistically significant improvement in time devoted to unpaid care work by men.  Once accounted for other confounding covariates, women employed as regular workers devote around two hours in unpaid household work per day less than women who are not employed. Similarly, women employed as regular or casual workers spend close to 16 minutes less in unpaid care work than women out of the workforce. Thus, in the case of women, employment participation results in a reduction in the amount of time they devote to unpaid household and care work at home. 
\includepdf[page={1,2}]{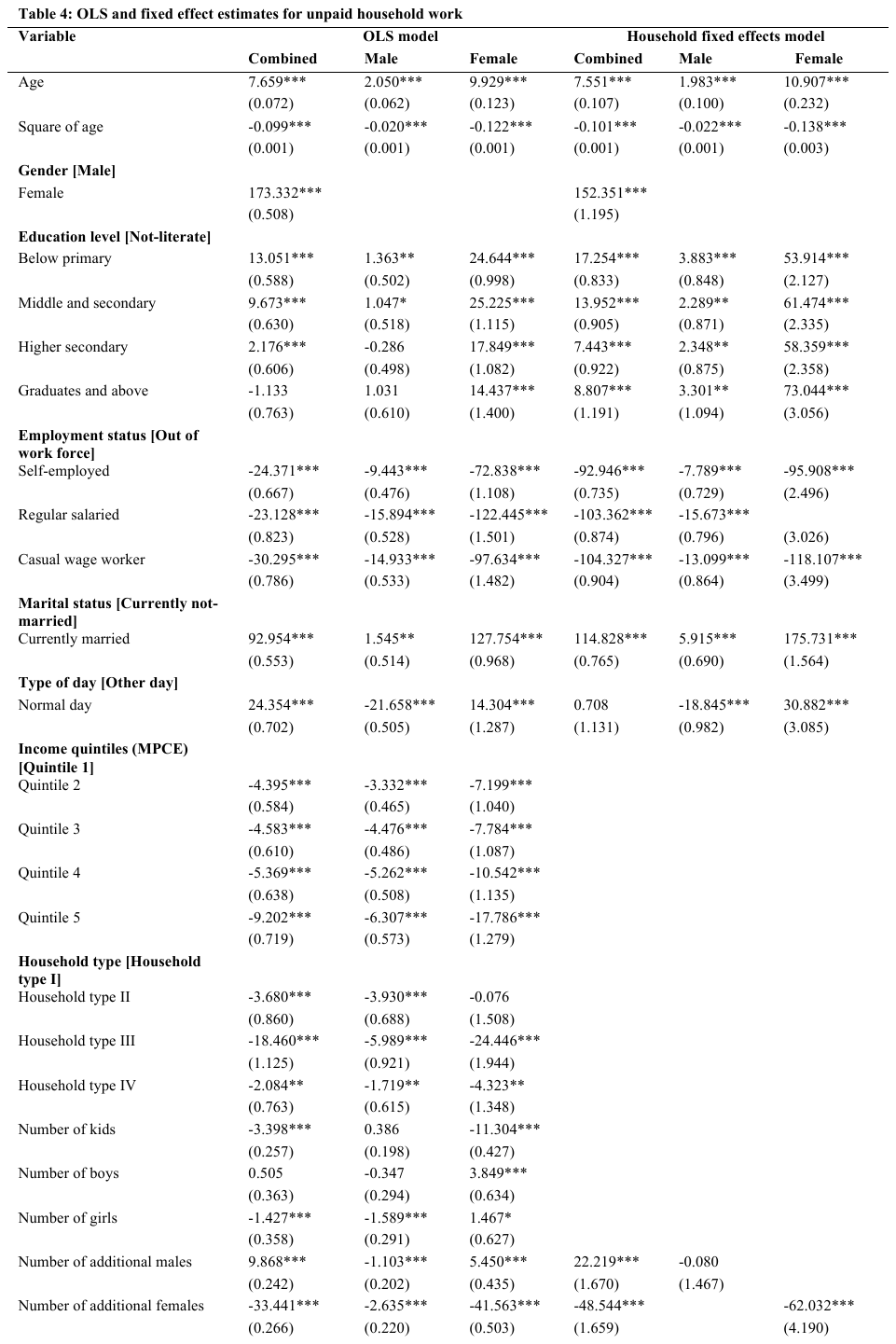}
In order to capture the relationship between individual's education and the time spent by them on unpaid household and care work , we have introduced educational attainment as a categorical variable in the regression model with not-literates as the reference category. The regression coefficient shows that women with primary education and middle school education spend 25 minutes more than the reference group. For women with higher-secondary education, the amount of time spent on unpaid household work increases by 18 minutes compared to not-literates and women in the highest educated group (graduate and above) spend 14 minutes more than women who are not-literate. After accounting for other covariates, education-based differences in the amount of time spent in unpaid care work are small for women. To summarise, with an increase in education, women's time spent on unpaid work did not decrease, whereas men continued to contribute negligibly to unpaid household work irrespective of their educational attainment. Notably, \cite{amarante2018unfolding} in their study of Latin American countries also find that women's response to unpaid household work with respect to improvements in educational attainments is more evident as compared to men.

We have incorporated both age and age-square as explanatory variables in our regression models to account for the non-linear relationship between age and the amount of time spent in unpaid household work. The coefficients of age and age-square show that for both women and men, the amount of time devoted in unpaid work varies with age in an inverted U-shaped pattern. Women spend the maximum amount of time on unpaid household work when they reach 40 years of age, and the corresponding age for men is 50 years. We have categorised all households into four different groups: single generation household, two-generation household, three-generation household, and others based on age of individual's within households (see Table 2). After controlling for other  factors, women in three-generation households spend 24 minutes  less on unpaid household work when compared to women in the reference category, that is, the one-generation households. This can be possibly attributed to sharing of unpaid household work by women in three-generation households. On the contrary, women in three-generation households spend 40 minutes more on unpaid care than women in one-generation households. This may be owing to additional members needing care in multiple generation households.

We categorise children by their age group as the amount of care needed varies with age and younger children may require more care from their parents than older children. As expected, one additional child in the age group 0-5 years in the household increased women's unpaid care work by 39 minutes. The corresponding increase for males is merely 10 minutes. It clearly shows that, women bear the increased burden of care work due to additional children in the household. Presence of an additional female in the age group, 15-59 in the household, reduces women's time devoted to unpaid household and care work. The time spent by women on household work falls by 42 minutes. Thus, extra females within the household point to collaborative behaviour among adult women who often share unpaid household work. On the contrary, with an additional male (in the age group 15-59) in the household, women's unpaid household and care work increases slightly by five minutes. With the additional adult members' presence in the household, the time spent by men in unpaid household work reduces marginally. The presence of extra elderly females shows a negative and significant association with women's time spent in household work in a day. 
\includepdf[page={1,2}]{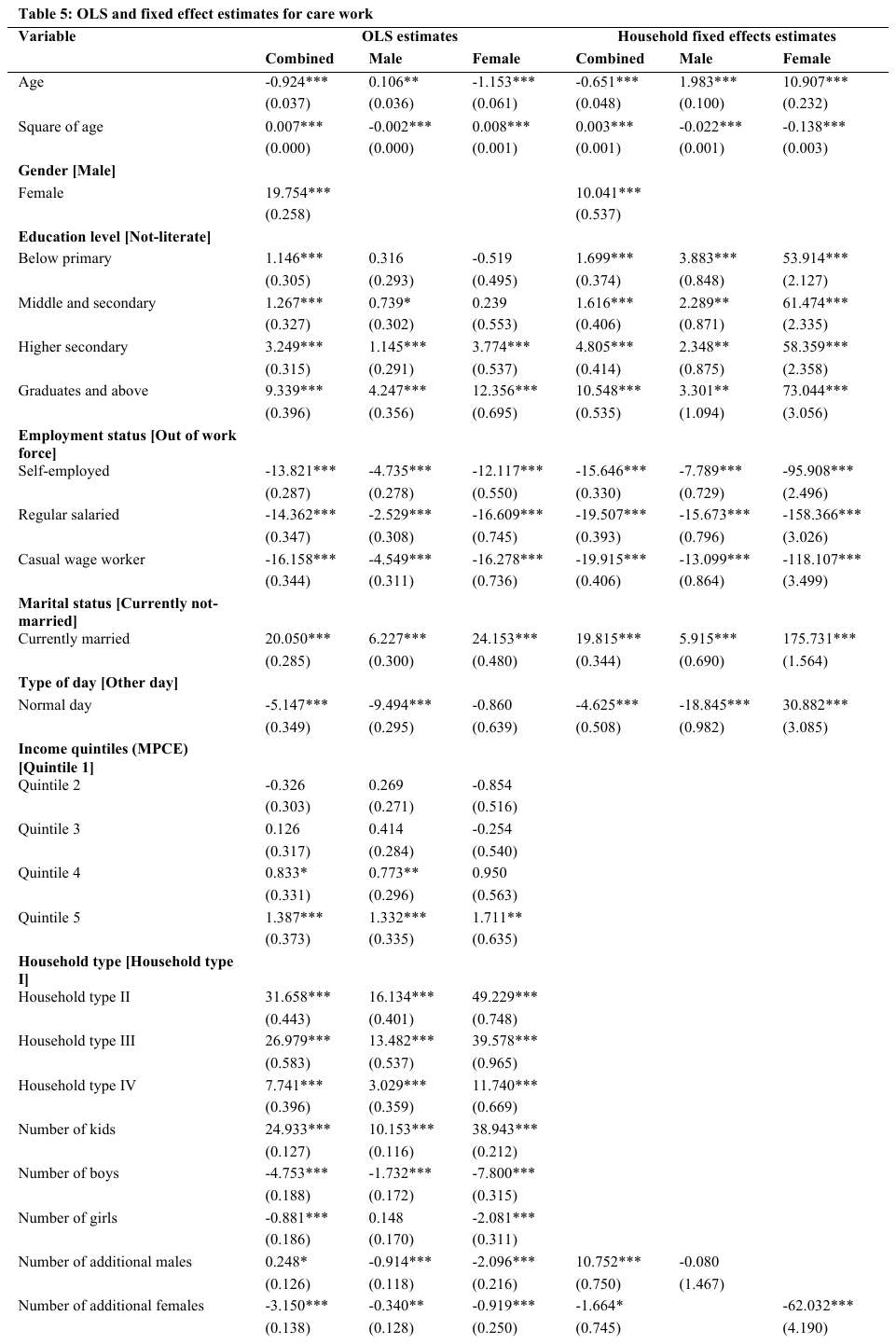}
Given the substantial rural and urban gaps in socioeconomic development, level of modernisation, and spread of education, an individual's place residence should be an essential predictor of time devoted to unpaid household work. However, keeping other factors fixed, the specification for both men and women shows that place of residence (rural or urban) has a negligible impact on the amount of time spent by women in unpaid household work per day.  Caste and religion are important markers of socioeconomic and cultural differences in India with differing labour force participation rate for women belonging to different caste and religious communities. The differences in gender norms as well as labour force participation among the different communities should be reflected in the time spent by women on unpaid household and care work. Nonetheless, we find that the relationship between caste and the amount of time spent on unpaid work at home is non-significant for both males as well as females. Similarly, differences in the amount of time spent by men and women belonging to various religions are statistically insignificant. In other words, caste and religion-specific cultures do not affect the time spent by women and men in unpaid household and care work in India. 
           
We have incorporated the monthly per capita expenditure (MPCE) as a proxy for household income in our regression model.  Women belonging to the fifth quintile, the top twenty per cent of income, devoted 17 minutes less to unpaid household work than the women from our reference category (the first quintile, the bottom twenty per cent). Outsourcing of routine unpaid household work such as washing clothes, sweeping floor and cooking, substantially reduces the time devoted by women to unpaid household work. For instance, a woman spends 51 minutes in unpaid household work in the household where the floor is swept manually compared to women in households where sweeping is outsourced.  The outsourcing of routine household work is more significant for women than for men. Household infrastructure such as washing machine and LPG stove are crucial predictors of the amount of time spent in unpaid household work by women in India. Women with access to these durable goods in their households tend to spend less time on unpaid household work in India.  
 
The specification including household fixed effect shows a slightly lower gender difference in unpaid household work and care work per day than other specifications.  However, the overall pattern of household fixed effect specification estimates in Panel A of Table 4 and Table 5 is similar to Panel  B in Table 4 and Table 5, and the results do not alter much.

\subsection{Decomposition analysis}

The regression results in the previous section show a considerable gender gap in the amount of time spent in unpaid household and care work per day. We employ Oaxaca-Blinder decomposition to understand the gap in the amount of time spent on the unpaid household and care work between women and men (Table 6). The decomposition method divides the average gender gap in unpaid work into two parts ‘explained’ and ‘unexplained’.  The ‘explained’ part can be attributed to differences in the average socioeconomic attributes between men and women. In other words, if women and men have the same level of socioeconomic covariates, then the total gap in unpaid work between women and men would reduce by the percentage of the ‘explained’ part. On the other hand, the “unexplained” part is attributed to unobserved factors. Therefore, in applying the decomposition method to the gender gap in unpaid work, the ‘unexplained’ gap can constitute the unobserved gender norms around unpaid household and care work.

 Column 2 in Table 6 summarizes results for model specification for gender gaps in unpaid household work. The differences in socioeconomic and demographic characteristics  between women and men only contribute to 27.5 percent of the gender gap in unpaid household work.  The differences in the socioeconomic and demographic factors between males and females can not explain the majority of the gap (72.5 percent) in the unpaid household between them. The decomposition estimates for the gender gap in unpaid care work (see column 3 in Table 6) parallels the decomposition results of the gender gap in unpaid work.  The unexplained gender gap in unpaid household work may be attributable to the prevailing  social norms that govern gender roles in Indian society, where women are expected to shoulder most of unpaid household and care work.
 \includepdf[page={1}]{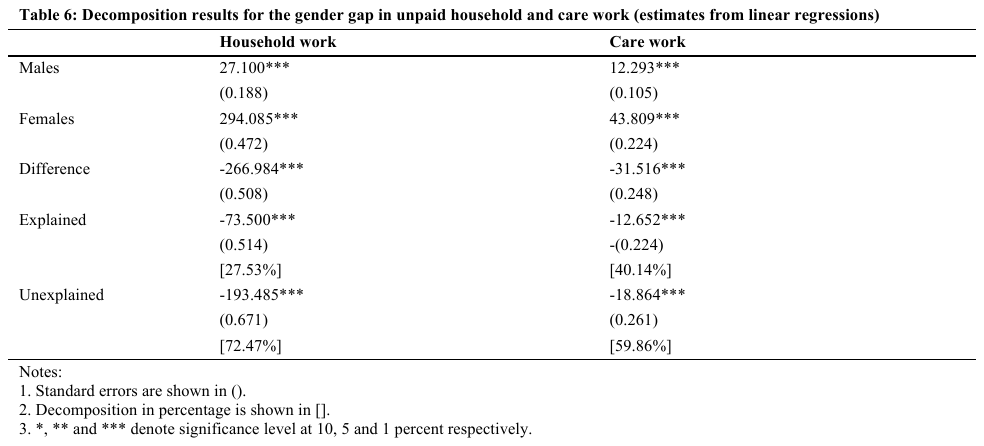}
 \subsection{Robustness Check }

To assess the robustness of the results of the gender gap in time spent in unpaid work, we estimated different alternative regression models. As discussed earlier, there is no consensus on the suitability of OLS or Tobit estimates to capture the association between the amount of time devoted to unpaid work per day and different socioeconomic and demographic covariates \citep{stewart2013tobit, foster2013tobit}. Thus, in Table 7 we present Tobit estimates. The set of control variables used in Tobit specification are same as that in OLS model. The estimates from Tobit specification in Table 7 are in the same direction as the OLS results presented in Table 4. For instance, in specification for men and women (see column 2 and column 5 in Table 7) after accounting for all other confounding covariates, the gender coefficient indicates that women spend about four and half hours and around one hour more than men in unpaid household work and unpaid care work per day respectively. Similarly, male and female individual Tobit estimates in Table 7 also show that the association between different covariates and the amount of time devoted to unpaid household and care work remained unchanged. For instance, married women spend around two hours more than not married women in unpaid household work. 

The decision to devote more time to unpaid household and care work involves a trade-off with time available for paid work and non-work activity in a day. Thus, to account for the interdependence of the four activities(i.e., unpaid household work,  unpaid care work, paid work and non-work activity), we estimated SUR model with the same set of covariates as OLS model. The purpose of using SUR model is to examine whether an increase in the time devoted to one type of activity  leads to a fall in the time spent in another kind of activity.

Table 8 presents SUR estimates of the four activities for the pooled sample of males and females.  The estimate of the female coefficient shows that women spend around three hours (186 minutes) more in unpaid household work and 19 minutes in unpaid care work more than men per day. On the contrary, women on average spend only 69 minutes less in paid work. Moreover, accounting for other factors, per day, women spend two hours less than men in non-work activities such as leisure and self-care. Working both within the home in unpaid household and care work coupled with the responsibilities of paid work reduces the time available for self-care and leisure considerably more for women as compared to men. 

The SUR specifications in Table 8, which examines how differences in time allocation vary with individual demographic and socioeconomic characteristics for men and women separately, confirms findings from OLS results on time devoted to unpaid household and care work. For example, being married and having children below five years in the household increases women's time allocation to unpaid household and care work. On the contrary, participation in the workforce, higher household income and additional adult females in the household reduce women's unpaid household and care work. The SUR estimates for men and women illustrate the higher burden of the total work time (i.e., time devoted to paid and unpaid care work) on women as compared to men, which in turn reduces women's available time for self-care and leisure. (see the last column of Table 8).

To check the robustness of our linear Oaxaca-Blinder decomposition, we also carried out Tobit decomposition. Qualitatively, both Oaxaca-Blinder and Tobit decomposition methods yield similar results. Both the methods show that differences in covariates between men and women cannot explain most of the gender gap in unpaid household work and care work. The results are presented in Table 9. The differences in the endowments between men and women can only explain  34 percent of the gap in time devoted to unpaid household work between them. The remaining 66 percent of the difference constitutes the unexplained part. The sizeable unexplained gender gap could be due to gender discrimination or other unobserved factors such as social norms. 

Women's bargaining power in the marital relationships within households is determined by her ability to enter the labour market and earn an income \citep{kan2008does}. Unfortunately wage data has not been collected in  the TUS-2019. To address this data gap, we have predicted wages using another nationally representative data set, the Periodic Labour Force Survey 2018-19. Further, to check the robustness of OLS results and decomposition analysis, we have repeated the analysis by incorporating the predicted wage variable in our regression and decomposition models. Regardless of adding log-wages as additional covariates in our analysis, the OLS and Tobit estimates remain largely unchanged (see Table A4). For example, specification 1 in column 2 in Table A4 shows that after accounting for log-wages and other covariates, women work around three hours more than men in unpaid household work. Similarly, Table A5 presents SUR results after controlling for wages. The SUR estimate are similar to OLS estimates. The decomposition results follow suit. The Oaxaca-Blinder decomposition estimates show that the majority of the gap in time devoted to unpaid work still cannot be explained by differences in covariates between women and men despite incorporating wage into the analysis (see Table A6 and Table A7). 

\includepdf[page={1,2}]{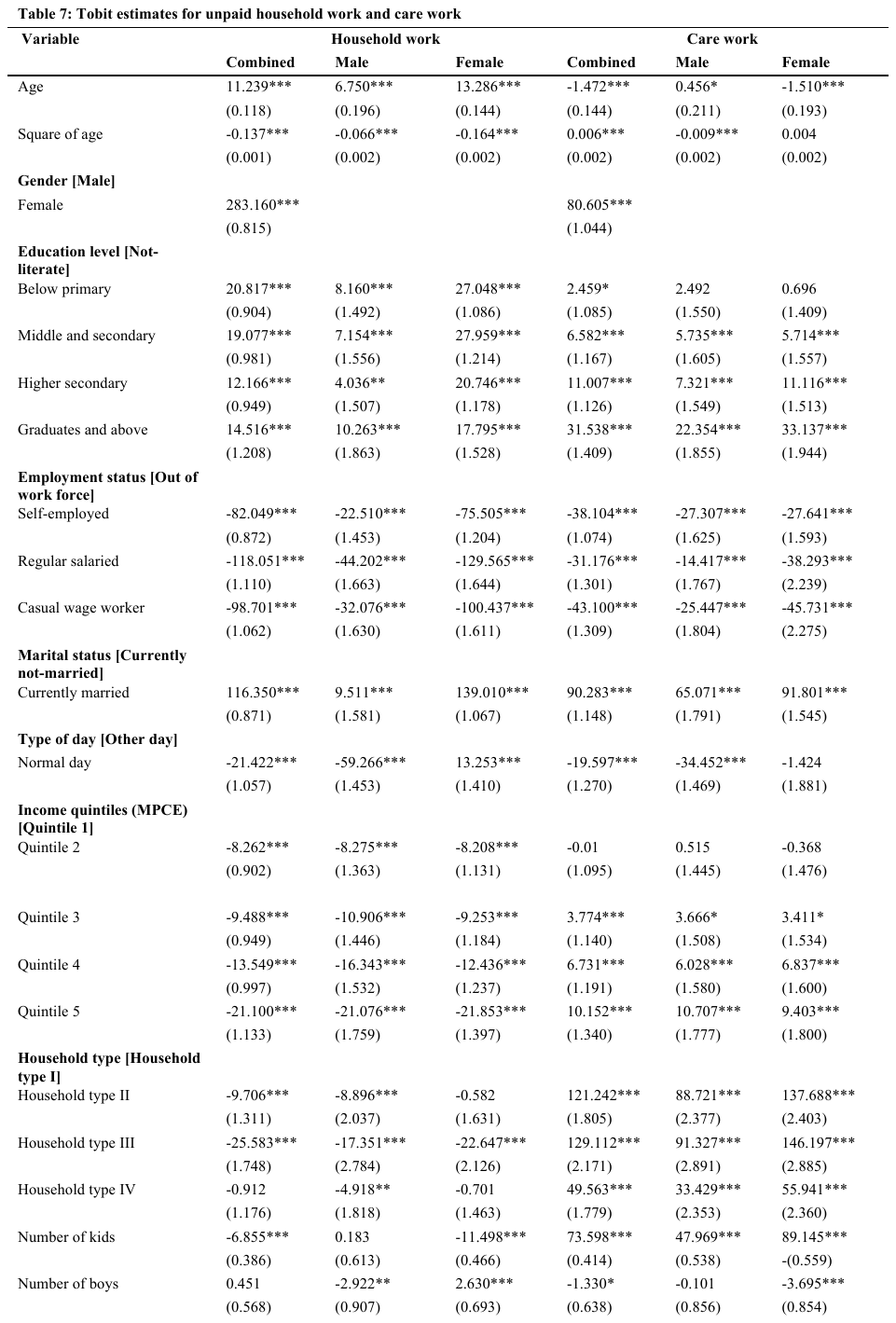}
 
\includepdf[page={1,2}]{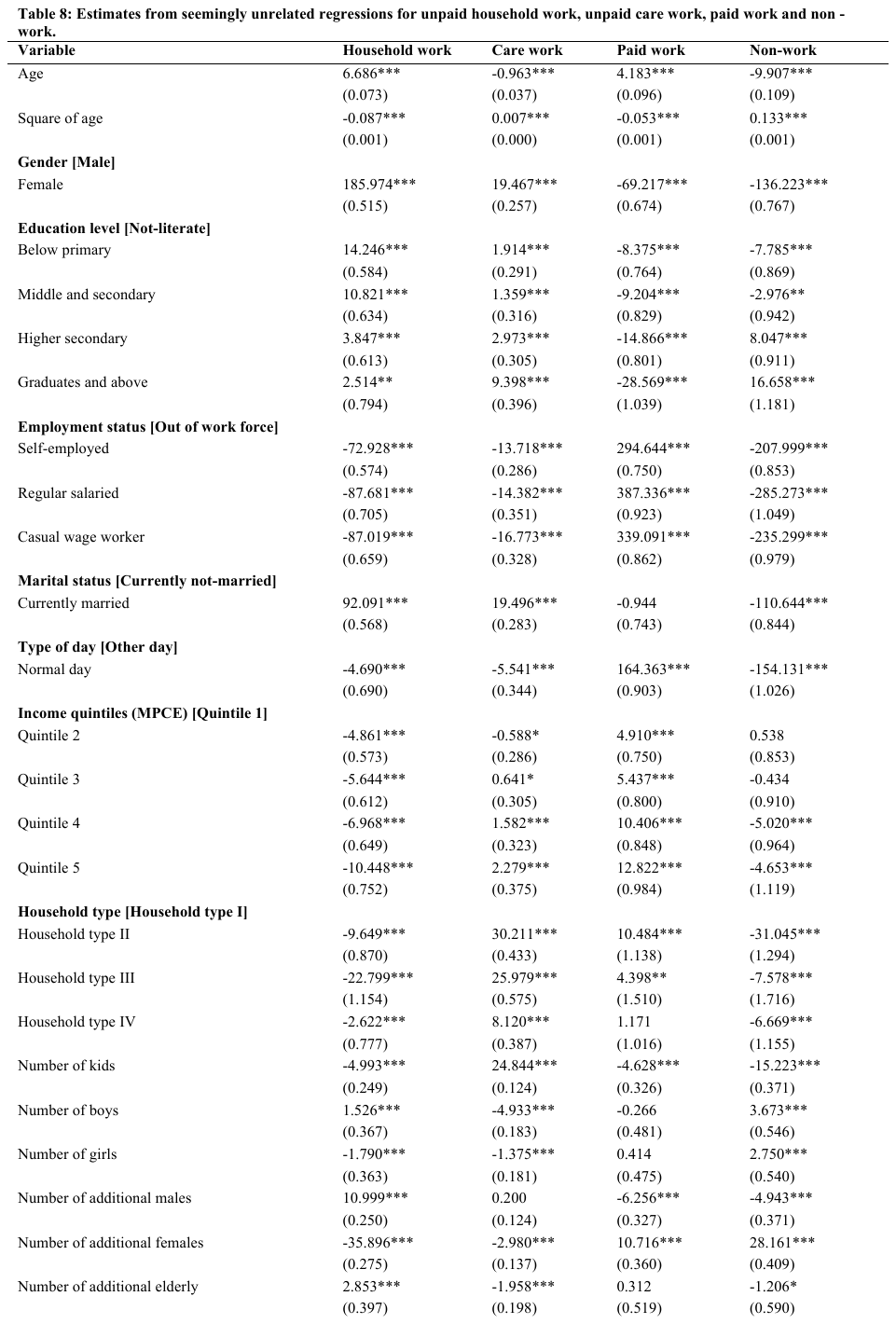} 

\includepdf[page={1}]{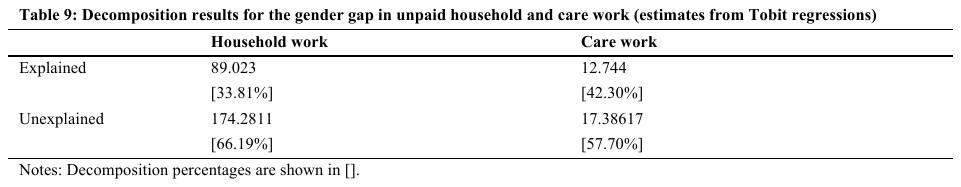} 

\section{ Conclusions} \label{sec:Conclusions}

Several studies have highlighted the gender gap in unpaid work in developed and developing countries. Nevertheless, due to lack of nationally representative data in India, only a limited number of studies have examined the gender gap in unpaid household and care work. This study using TUS 2019 examines the gap in time spent in unpaid household and care work between women and men in India.  
 
The results show that women spend disproportionately more time in unpaid work as compared to men in India. For example, our results show that after accounting for individual characteristics and household factors, women spend around  three hours (173 minutes)  more  in unpaid household work per day (see column 2 in Table 2).  The OLS estimates demonstrate that married women devoted more time to unpaid housework and unpaid care work than unmarried women. In addition, women's participation in paid work, the presence of additional females in the household, and possession of time-saving infrastructure for cooking, cleaning, and washing clothes are significant predictors of women's time devoted to unpaid household activity. Variation in time dedicated to unpaid household work by women with an increase in education and household per capita income is very small. Similarly, caste and religion of women has negligible effect on time spent in unpaid work after controlling for other relevant covariates in the model. Contrary to expectation, men with an increase in education do not contribute substantially to unpaid household work. The SUR results highlight that though women spend around one hour less time in paid work than men, their allocation of time to total (both paid and unpaid) work is higher than that of men. In other words, women's higher time allocation to unpaid household and care work leaves them with much lower time available for their non-work activities such as leisure and self-care than men. In the context of China, \cite{dong2015gender} similarly found that in addition to paid work  the burden unpaid at home work  left  women more time poor than men. However,  it is essential to note that the regression results only bring out the statistical association that cannot be interpreted causally due to potential endogeneity issues.  
 
The regression decomposition results show that difference in socioeconomic and demographic factors between males and females only contribute a small part of the average gender gap in unpaid household  and care work. That is, most of the gap in time spent in unpaid work between males and females is unexplained. The unexplained gap may include unobserved gender-based social norms that pin the responsibility of unpaid household and care work on women. Most of the gap in time devoted to unpaid household and care work remains unexplained despite introducing predicted log-wages as an additional variable in decomposition analysis. 
 
The sizeable unexplained gap in time allocation to unpaid work between women and men may indicate the persistence of gendered attitudes and behaviour towards unpaid household and care work in the Indian society. Our results are aligned with findings from Latin America \citep{amarante2018unfolding}  and other developing countries such as Kyrgyzstan \citep{kolpashnikova2020gender} and Iran \citep{torabi2020spouses} that the majority of the gender gap in unpaid work cannot be explained by differences in the socio-economic characteristics of men and women. Increasing literacy rates and women's education over the last few decades in India does not seem to translate into attitudinal change in norms and perception of gender roles. As per National Family and Health Survey 1998 and 2016,  more than 30 percent of women and men reported that men beating their wives was justified if women neglect housework and children. Similarly, \cite{thorat2020persisting}, using Social Attitudes Research, India  (SARI) 2016-18  data show that across caste groups, around 40 percent of men and women reported that women should not participate in labour force if the husband earns enough.  Evidence from NFHS and SARI (2016-18) data indicates a high prevalence of gender division in outside work and household work, signalling the continued prevalence of the social norm that unpaid household and care work is the primary responsibility of women in India. In order to achieve equitable distribution of unpaid work between women and men, there is a need to provide more opportunities for women to participate in paid employment. Further policy level initiatives such as providing access to time-saving infrastructure and strengthening free early child care support programs will help to reduce the burden of unpaid work for women in India.

\section{Acknowledgements} \label{sec:Acknowledgements}
The authors are grateful to Sunny Jose for his invaluable comments and insights.

\clearpage
\singlespacing
\setlength\bibsep{0pt}
\bibliographystyle{apacite}

\bibliography{timeuse.bib}

\includepdf[page={1,2}]{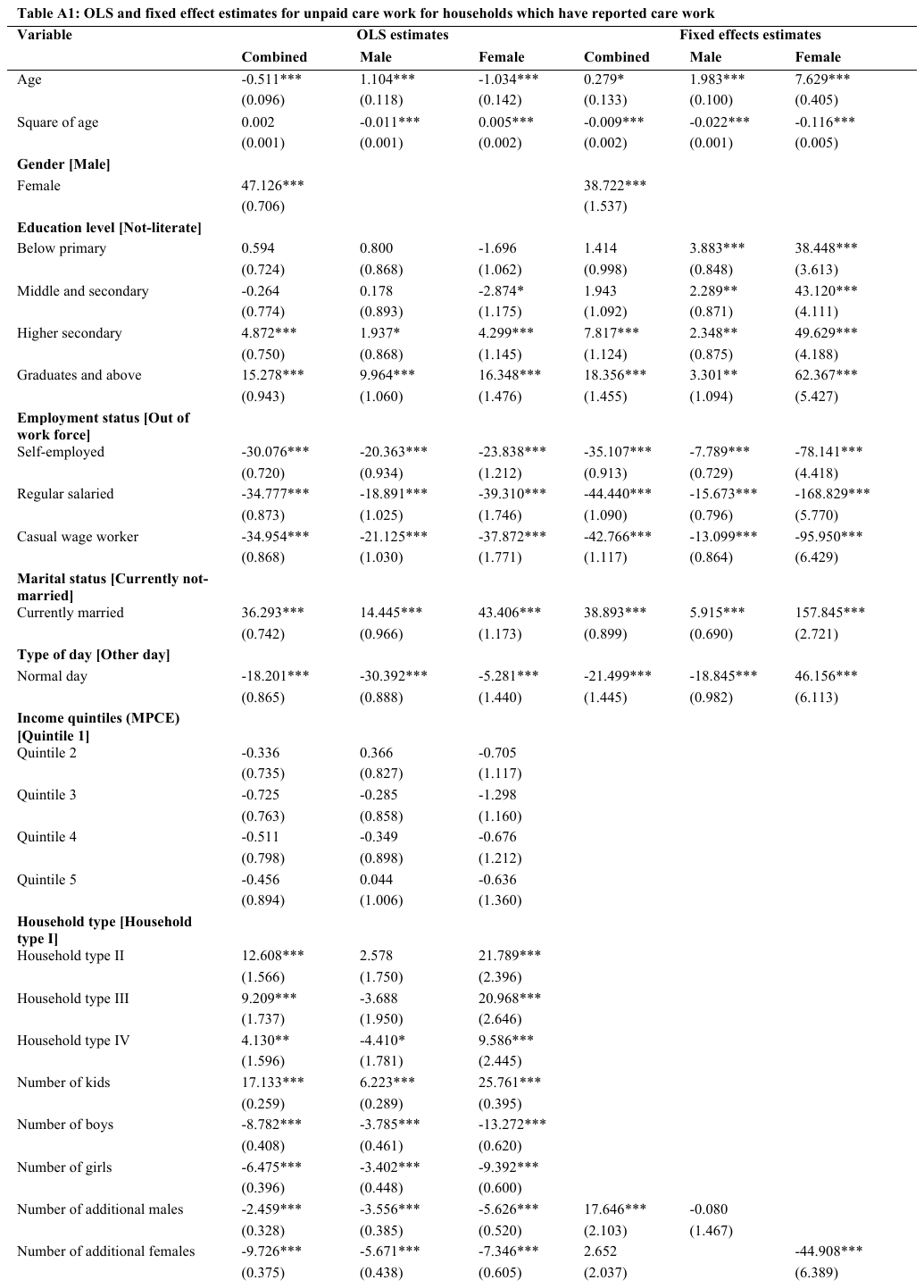}
\includepdf[page={1,2,3}]{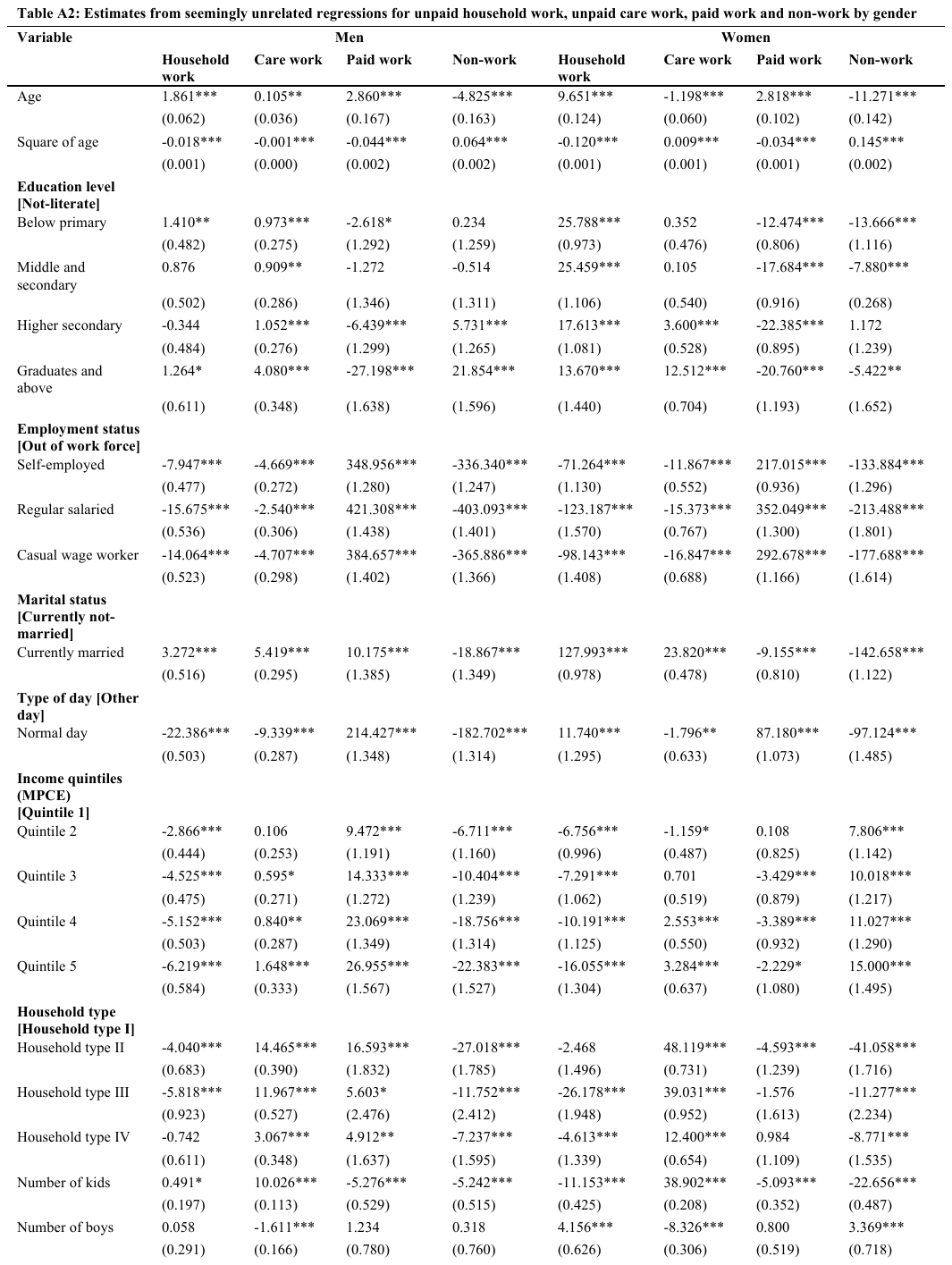}
\includepdf[page={1}]{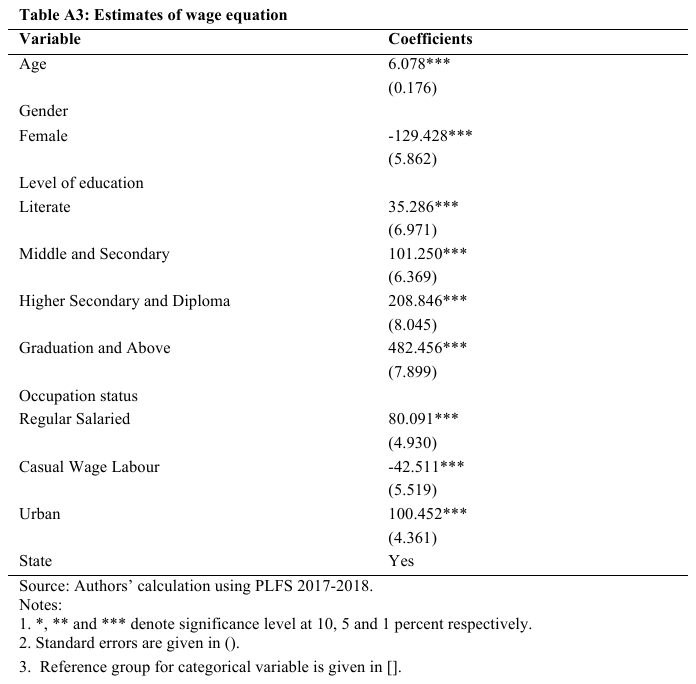}
\includepdf[page={1,2}]{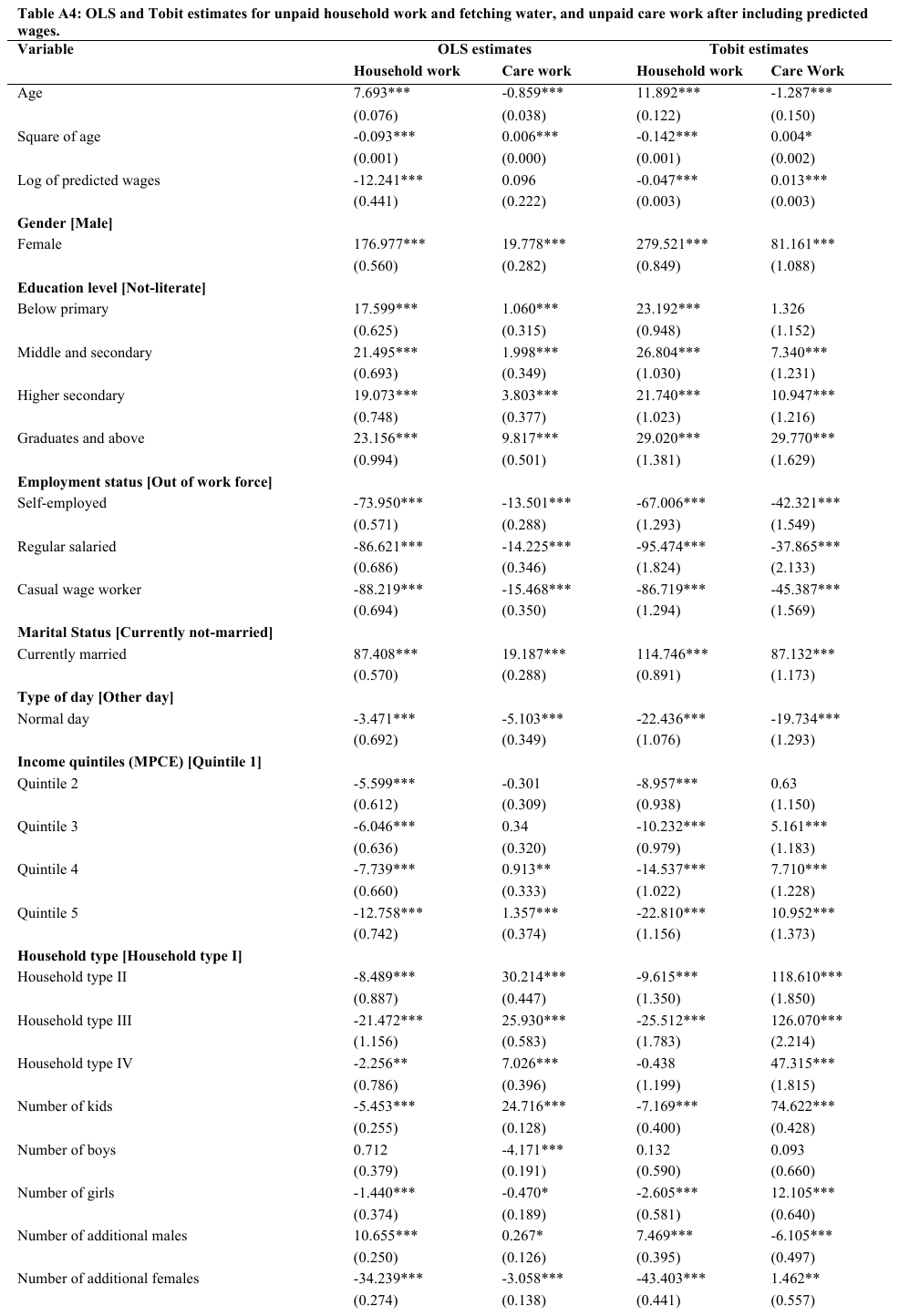}
\includepdf[page={1,2}]{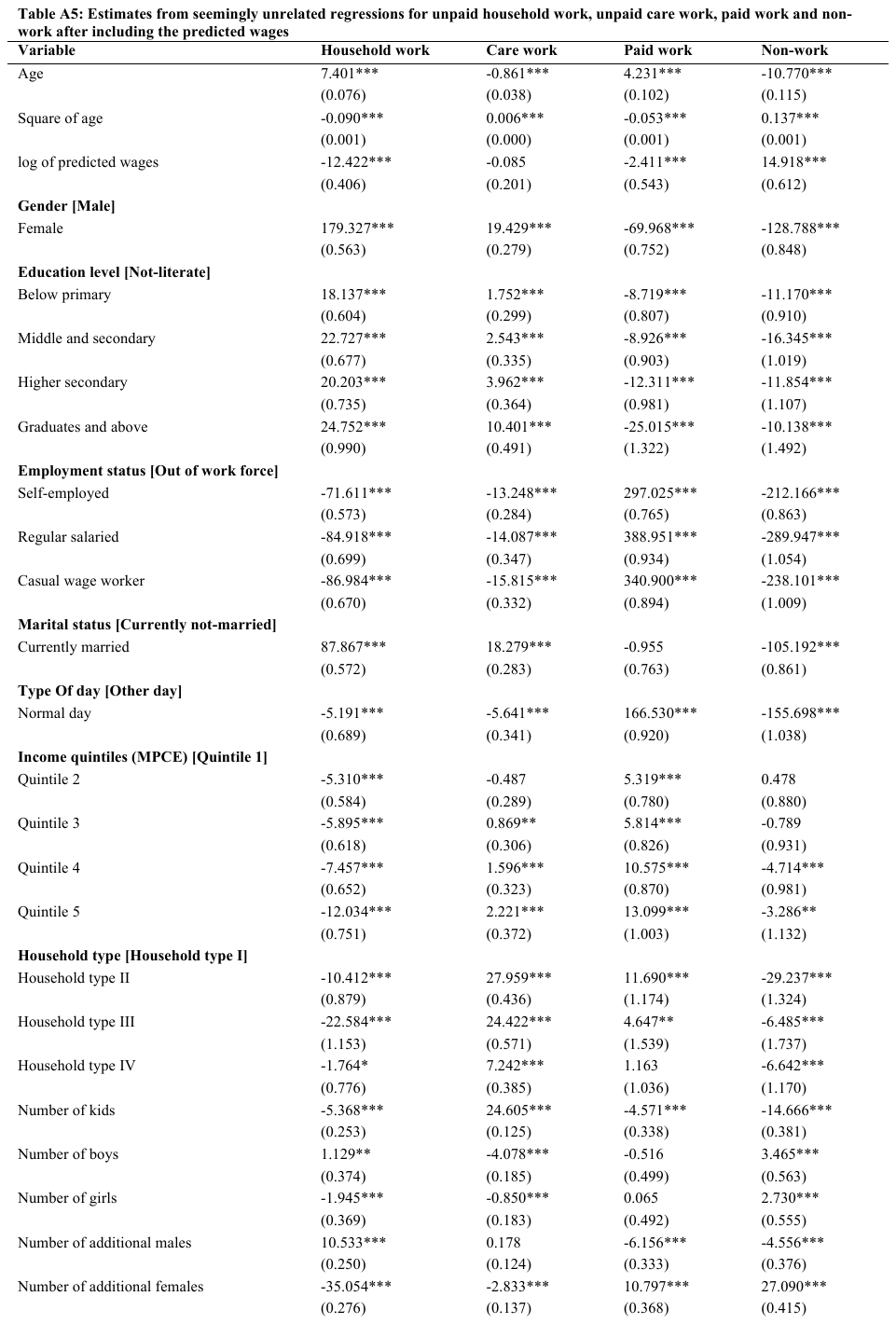}
\includepdf[page={1}]{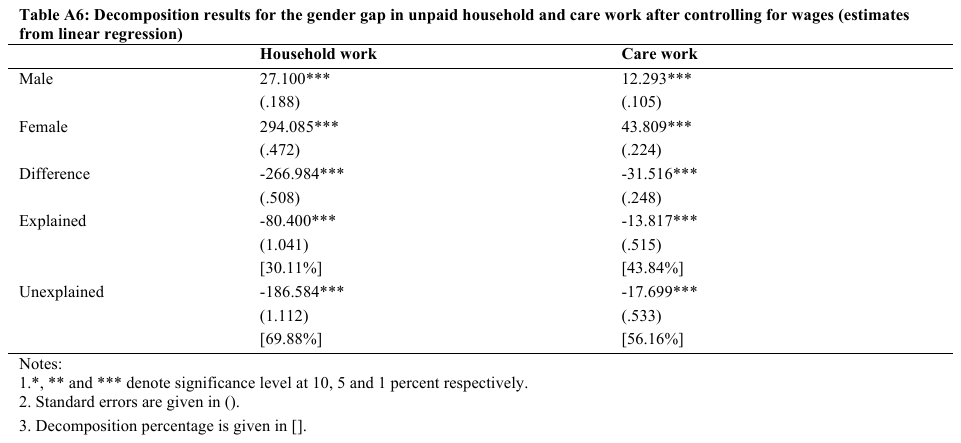}
\includepdf[page={1}]{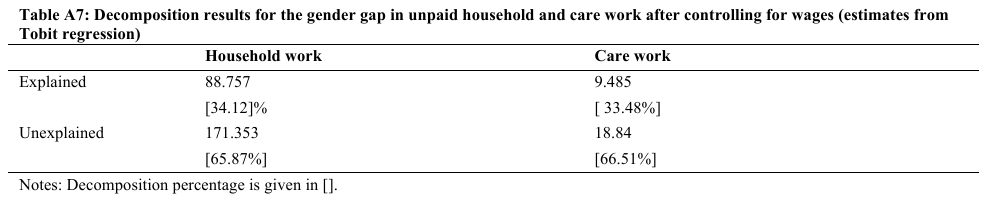}

\clearpage
\newpage

\end{document}